\documentclass[prl,twocolumn,superscriptaddress,showpacs]{revtex4}
\usepackage{epsfig}

\def\vec#1{\hbox{\boldmath $#1$}}

\begin{document}
\title{Indirect dissociative recombination of LiH$^+$ molecules 
fueled by complex resonance manifolds}
\author{R. \v{C}ur\'{\i}k}
\email[Corresponding author: ]{curik@colorado.edu}
\affiliation{Department of Physics and JILA, University of Colorado, Boulder, CO
80309-0440}
\author{Chris H. Greene}
\affiliation{Department of Physics and JILA, University of Colorado, Boulder, CO
80309-0440}

\begin{abstract}
The LiH$^{+}$ molecule is prototypical of the {\it indirect} 
dissociative recombination (DR) process, in which a colliding electron
destroys the  molecule through Rydberg capture pathways. This Letter
develops the first quantitative test of the Siegert state
multichannel quantum defect theory description of indirect DR
 for a diatomic molecular ion. The R-matrix approach is adopted to
calculate \textit{ab-initio} quantum defects, functions of the internuclear
distance that characterize both Rydberg states and the zero-energy
collisions of electrons with LiH$^{+}$ ions. 
The calculated DR rate coefficient 
agrees accurately with recent experimental data \cite{Zajfman-LiH-2001}. We
identify the doorways to fast indirect DR as complex resonance manifolds, 
which couple closed channels having both high and low principal quantum numbers.  
This sheds new light on the competition between direct and
indirect DR pathways, and suggests the reason why previous theory
 underestimated the DR\ rate by an order of magnitude.
\end{abstract}

\maketitle

The LiH molecule was one of the first participants in early universe
chemistry, and it played a key role in the cooling of primordial gases. 
Therefore, the relative abundance of LiH and its formation process have drawn
significant attention in models of the universe. Stancil \textit{et
al.} \cite{Stancil-Dalgarno-1996} noticed that direct radiative association
of neutral atoms, 
$
\mathrm{Li+H}\longrightarrow \mathrm{LiH}+\nu  
$
\ occurs at the very slow rate coefficient of 
$\sim 10^{-20}$ cm$^{3}$s$^{-1}$ . 

On the other hand, radiative association with ionic hydrogen 
$
\mathrm{Li+H}^{+}\longrightarrow \mathrm{LiH}^{+}+\nu  
$
\ is predicted to occur \cite{Dalgarno-Kirby-1996, Gianturco-Giorgi-1997} at
the much higher rate of $\sim 10^{-15}$ cm$^{3}$s$^{-1}$. The resulting LiH$%
^{+}$ and LiH abundances are controlled by photoionization, collisions, and by
dissociative recombination (DR) with free electrons. A recent DR experiment
for LiH$^{+}$ \cite{Zajfman-LiH-2001} has measured the DR rate
coefficient for collisions at $T=$139K (12 meV) to be 
(6$\pm $2) x 10$^{-7}$ cm$^{3}$s$^{-1}$. From
numerous \textit{ab-initio} calculations of the LiH and LiH$^{+}$ potential
surfaces \cite{Boutalib-Gadea-1992,Berriche-Gadea-1995,Gemperle-Gadea-1999,
Yiannopolou-Jeung-1999,Altunata-Field-2003,Florescu-Gadea-2004} there is no
Born-Oppenheimer neutral state that crosses the ionic ground state potential
curve, anywhere near the ionic minimum. Thus the high measured DR rate for
such an \textit{indirect} or \textit{non-crossing} process is challenging to
reconcile with existing theoretical results. The rate coefficient estimated
theoretically in \cite{Stancil-Dalgarno-1996} is 2.6 $\times $ 10$^{-8}$ cm$%
^{3}$s$^{-1}$. Another theoretical study by Florescu \textit{et al.} \cite%
{Florescu-Gadea-2004} applied multichannel quantum defect methods, with the
relevant nonadiabatic coupling elements obtained from a generalized
Hellmann-Feynman theorem \cite{Stolyarov-Child-2001} to calculate that the
139K DR rate coefficient should equal 3.6 $\times $ 10$^{-8}$ cm$^{3}$s$%
^{-1} $. Since both of these theoretical studies underestimate the DR rate
for this simple diatomic by more than an order of magnitude, it shows that
the proper physical description of the underlying mechanism for indirect DR
processes continues to challenge our theoretical understanding.

Quantum defect theory in connection with frame transformation into a basis
of Siegert vibrational states \cite{Siegert-1939, Tolstikhin-1998} has shown
promise in describing DR for a model diatomic \cite{Hamilton-Greene-2002}
and for the triatomic H$_{3}^{+}$ molecule \cite{Kokoouline-Greene-2003},
systems for which indirect Rydberg state pathways dominate. But to date
there has been no rigorous test of the Siegert-state-based MQDT formulation,
for an experimentally-studied system for which the relevant quantum defect
matrices have been determined directly in an \textit{ab initio}
scattering-type calculation. \ Accordingly, the main goal of the current paper
is to analyze dissociative collisions between a low-energy electron and the
LiH$^{+}$ ion, as a fundamental prototype system that provides a stringent
test of this combination of theoretical elements: R-matrix theory, MQDT, and
frame-transformation theory based on Siegert pseudostates \cite%
{Chang-Fano-1972, orange-review, Seaton, GreeneJungen}.

The body-fixed adiabatic eigenquantum defects $\mu _{\gamma }^{\Lambda }(R)$
are related to the energy differences (in a.u.) between the potential curves of the
ion $U^{+}(R)$ and the neutral Rydberg states $U_{n\gamma }^{\Lambda }(R)$
by Mulliken's equation%
\begin{equation}
U_{n\gamma }^{\Lambda }(R)=U^{+}(R)-\frac{1}{2\left[ \nu _{\gamma }^{\Lambda
}(R)\right] ^{2}}\;,  \label{eq2-mulliken}
\end{equation}%
where the effective quantum number $\nu _{\gamma }^{\Lambda }(R)=n-\mu
_{\gamma }^{\Lambda }(R)$ and $\Lambda $ denotes the projection of the
Rydberg electron angular momentum $l$ onto the axis of the diatomic
molecule. Of course $l$ is not a good quantum number,
so $\gamma $ is an eigen-index distinguishing different Rydberg series
of LiH. Fundamentally, body-frame quantum defects are represented by a
matrix $\mu _{ll^{\prime }}^{\Lambda }(R)$ with different partial-waves
coupled by off-diagonal elements, after which the $\mu _{\gamma }^{\Lambda
}(R)$ are obtained as its eigenvalues. \textit{Ab-initio} potential energy
surfaces by themselves provide no direct information about off-diagonal
couplings, or equivalently, about the eigenvectors $U_{l\gamma }(R)$, and
many authors tend to neglect them \cite{Florescu-Gadea-2004} or estimate
them using a two-channel Landau-Zener curve-crossing formula \cite%
{Landau-Zener} or sometimes by fitting them in a diabatic representation 
\cite{Tonzani-HCO}.

But the present study exploits the familiar MQDT theorem \cite{Seaton}
that smoothly connects quantum defects (multiplied by $\pi $) at energies
just below the ionization threshold to $R$-dependent short-range scattering
phase shifts (or multichannel scattering or reaction matrices) just above
the threshold: 
\begin{equation}
\pi \;\mu _{ll^{\prime }}^{\Lambda }(R)=\sum_{\gamma }\;U_{l\gamma
}(R)\;\delta _{\gamma }^{\Lambda }(R)\;U_{l^{\prime }\gamma }(R)\;.
\label{eq2-QD-matrix}
\end{equation}%
Here $\delta _{\gamma }^{\Lambda }$ are the low- or zero-energy eigenphases
for e$^{-}$ + LiH$^{+}$ collision and the eigenvector matrix is $\underline{%
U}(R)$ which transforms the short-range $K$-matrix into diagonal
form. 
\begin{figure}[tbph]
\begin{center}
\epsfig{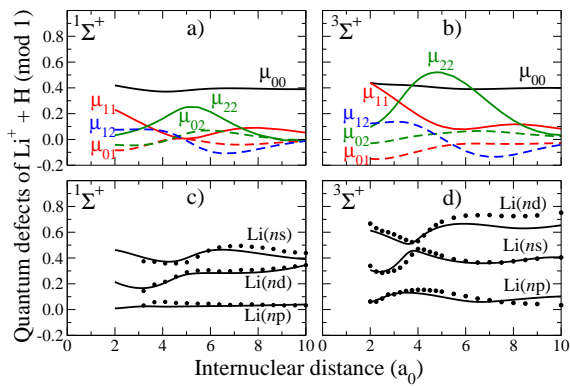}
\end{center}
\vspace{-2mm}
\caption{ Upper panels: The calculated quantum-defect matrix
elements are shown versus the internuclear distance - panel a) singlets, panel b)
triplets. Lower panels: The solid curves are the eigenvalues of the matrices in a)
and b). The black dots are eigenquantum defects extracted from n=4
states in extensive bound state CI calculations \protect\cite{Florescu-Gadea-2004}. }
\label{fig-sigma-qdm}
\end{figure}
We used the diatomic UK R-matrix package \cite{Morgan-Rmat-1997} to
calculate the short-range $K$-matrix by matching to coulomb
functions at an R-matrix boundary of $r_{0}=25a_{0}$. The target was
described by an augVTZ STO basis set \cite{Ema-Paldus-2002}. From the
calculated $^{1}\Sigma ^{+}$ and $^{3}\Sigma ^{+}$ quantum defects shown in
Fig.~\ref{fig-sigma-qdm}, the $s$-wave quantum defect is only
weakly perturbed from its Li$^{+}$ limit of 0.399, over a wide range of
internuclear distances. Thus coupling to nuclear motion will be controlled
by higher partial waves, namely $p$- and $d$-waves. The $^{1}\Pi $ and $%
^{3}\Pi $ quantum defects have been calculated to be an order of magnitude
smaller, and they have negligible impact on the final DR results.

The internuclear distance $R$ is a good body frame ``quantum number'' when all
electrons are confined within the box specified by $r_{0}$ and the
Born-Oppenheimer approximation is strictly valid. The vibrational frame
transformation connects $R$ with the laboratory-frame quantization of this
degree of freedom expressed by vibrational wave functions $\phi _{j v  }(R)$%
. Siegert pseudostates \cite{Siegert-1939,Tolstikhin-1998} provide a unified
description of the bound vibrational states and the vibrational continuum.
In the present case they solve the vibrational Schr\"{o}dinger 
equation and boundary conditions: 
\begin{eqnarray}
\left[ -\frac{d^{2}}{dR^{2}}+2MU^{+}(R)+\frac{j(j+1)}{R^{2}}-k_{j v  }^{2}%
\right] \phi _{j v  }(R)=0\;,  \label{eq3-Siegert-Hamiltonian}
\\
\phi _{j v  }(0)=0\;\;\;\;\;;\;\;\;\;\;\left. \left( \frac{d}{dR}-ik_{j v 
}\right) \phi _{j v  }(R)\right\vert _{R_0}=0\;.
\label{eq3-Siegert-Boundary}
\end{eqnarray}
In the above equations $j$ is a rotational quantum number of the ion, $M$
stands for its reduced mass, while $R_0$ denotes a nuclear radius beyond
which we approximate the interaction potential in (\ref%
{eq3-Siegert-Hamiltonian}) to be constant and $\phi _{j v  }(R)=\exp
(ik_{j v}R)$ for $R\geq R_0$. 
\begin{figure}[tbph]
\begin{center}
\epsfig{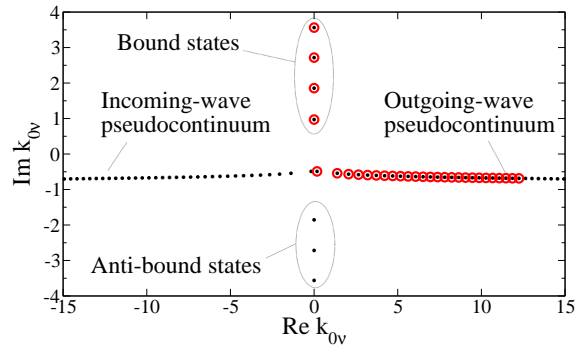}
\end{center}
\vspace{-2mm}
\caption{ Distribution of the $j=0$ Siegert momentum 
eigenvalues from Eqs.(\protect\ref%
{eq3-Siegert-Hamiltonian},\protect\ref{eq3-Siegert-Boundary}) in the
complex plane. The circled states are the ones 
used in the calculations of this paper. 
}
\label{fig-cplane}
\end{figure}
Fig.~\ref{fig-cplane} shows an example of the $j=0$ Siegert 
state momentum eigenvalue distribution. Because the nuclei are confined
within $R_0=10a_{0}$ we only obtain 4 bound states, in contrast with
the expected total of 7 bound states found in \cite%
{Berriche-Gadea-1995,Florescu-Gadea-2004}. \ However for the lower
vibrational levels that fit inside, agreement in the level
spacing is achieved within 2 cm$^{-1}$.

Because the orthogonality relation between two different
Siegert pseudostates is slightly modified \cite{Tolstikhin-1998}, 
a surface term \cite%
{Hamilton-Greene-2002} is added to the standard frame transformation
integral \cite{Greene-Jungen-1085} yielding 
\begin{eqnarray}
S_{l v  ,l^{\prime } v^{\prime }}^{\Lambda }(j,j^{\prime })
&=&\int_{0}^{R_0}dR\;\phi _{j v  }(R)\left( e^{2i\pi \underline{\mu }%
^{\Lambda }(R)}\right) _{l,l^{\prime }}\phi _{j^{\prime } v^{\prime }}(R) 
\nonumber \\
&+&i\frac{\phi _{j v  }(R_0)\left( e^{2i\pi \underline{\mu }^{\Lambda
}(R)}\right) _{l,l^{\prime }}\,\,\phi _{j^{\prime } v^{\prime }}(R_0)}{%
k_{j v  }+k_{j^{\prime } v  ^{\prime }}}\;.
\end{eqnarray}%
The underline in this equation denotes that $\underline{\mu}%
^{\Lambda}$ is a matrix with indices $\mu _{ll^{\prime }}^{\Lambda }(R)$.
Moreover, the rotational indices $(j,j^{\prime })$ of this body-frame
$\underline{S}$-matrix do not give rotational transition probabilities
- they only serve as a reminder that the vibrational functions exhibit a $j$%
-dependence through the centrifugal term in Eq.(\ref{eq3-Siegert-Hamiltonian}%
).
To reiterate, the rotational frame transformation transforms a 
set of body-frame quantum numbers $(l,\Lambda ,J)$ into a set of 
laboratory-frame quantum numbers $(l,j,J)$. The total angular momentum $J$ 
is defined via $\vec{J}=\vec{l}+\vec{j}$. The LiH$^{+}$ ion is treated in Hund's 
case (b), with spin-orbit coupling neglected. Definite total
parity $\eta =(-1)^{l+j}$ is also enforced, whereby the
short-range laboratory-frame scattering matrix is:   
\begin{equation}
S_{lj v  ,l^{\prime }j^{\prime } v  ^{\prime }}^{J\eta }=\sum_{\Lambda
}U_{Jjl}^{\Lambda \eta }\;S_{ v  l, v  ^{\prime }l^{\prime }}^{\Lambda
}(j,j^{\prime })\;U_{Jj^{\prime }l^{\prime }}^{\Lambda \eta }\;.
\label{eq4-Rotational-FT}
\end{equation}%
The real, orthogonal rotational transformation matrix $%
U_{Jjl}^{\Lambda \eta }$ is taken from \cite{Chang-Fano-1972} and will not be
repeated here.

As is familiar in MQDT applications, the \textquotedblleft
short-range\textquotedblright\ or \textquotedblleft
unphysical\textquotedblright\ scattering matrix $\underline{S}$ in
Eq.(\ref{eq4-Rotational-FT}) is diagonal in the $J$ and $\eta $ quantum numbers.
It represents an amplitude for electron-ion scattering from an initial
channel defined by $(l^{\prime }j^{\prime } v  ^{\prime })$ into a final
channel defined by $(lj v  ),$ but some of these channels in $\underline{S}$ are
typically closed energetically. The physically-relevant $%
S$-matrix is of course defined only in the open-channel space and is
obtained by the \textquotedblleft elimination of closed channels" formula 
\cite{orange-review}: 
\begin{equation}
\underline{S}^{\mathrm{phys}}=\underline{S}^{\mathrm{oo}}-%
\underline{S}^{\mathrm{oc}}\left[ \underline{S}^{\mathrm{cc}%
}-e^{-2i\underline{\beta}(E)}\right] ^{-1}%
\underline{S}^{\mathrm{co}}\;,
\label{eq5-Channels-Elim}
\end{equation}%
where the superscripts $\mathrm{o}$ and $\mathrm{c}$ respectively denote open 
and closed sub-blocks of the unphysical $S$-matrix (\ref%
{eq4-Rotational-FT}) and \underline{$\beta $}$(E)$ is a diagonal matrix of
effective Rydberg quantum numbers with respect to the closed-channels
thresholds: 
\begin{equation}
\beta _{ij}=\frac{\pi \delta _{ij}}{\sqrt{2(E_{i}-E)}}\;.  \label{eq5-beta}
\end{equation}%
The total energy of the electron + ion system is $E$, and $E_{i}$ is the
ionization threshold for channel $i\equiv \left( vjl\right)$.
Here as in Refs. \cite{Hamilton-Greene-2002,Kokoouline-Greene-2003} the high
ionization thresholds are described by a Siegert pseudo-continuum state with
finite widths, so $E_{i}$ and \underline{$\beta $}$(E)$ are {\it complex}.
This fact alone destroys the unitarity of $\underline{S}^{\mathrm{phys}}$ 
making it sub-unitary. The lost flux is associated with a trapped
Rydberg electron in a closed channel that represents a high-lying
vibrational state that is dissociative, with outgoing-wave character, and
has a complex vibrational energy and corresponding finite lifetime. The
departure from unitarity was identified by \cite%
{Hamilton-Greene-2002,Kokoouline-Greene-2003} as the dissociation
probability following electron impact in incident channel $i^{\prime }
$: 
\begin{equation}
\sigma _{i^{\prime }}^{J\eta }(\varepsilon _{i^{\prime }})=\frac{\pi }{%
2\varepsilon _{i^{\prime }}}\left[ 1-\sum_{i}S_{ii^{\prime }}^{\mathrm{phys}%
}(E)\;S_{i^{\prime }i}^{\dagger \,\mathrm{phys}}(E)\right] \;,
\label{eq5-DR-CS}
\end{equation}%
with the incident electron collision energy $\varepsilon _{i^{\prime
}}=E-E_{j^{\prime }v^{\prime }}$. This cross-section depends on the initial
channel $i^{\prime }=( v  ^{\prime }j^{\prime }l^{\prime });$ the collision
preserves the conserved quantum numbers $(J\eta )$. The experimentally
observable cross-section for dissociation following electron impact is then
\begin{equation}
\sigma _{j^{\prime } v  ^{\prime }}(\varepsilon _{j^{\prime }v^{\prime }})=%
\frac{1}{2j^{\prime }+1}\sum_{\eta Jl^{\prime }}(2J+1)\;\sigma _{l^{\prime
}j^{\prime } v  ^{\prime }}^{J\eta }(\varepsilon _{j^{\prime }v^{\prime
}})\;.  \label{eq5-CS-sum}
\end{equation}%
We
further average over a Boltzmann distribution of initial ro-vibrational
states of the ion at the temperature (T = 300K) appropriate to the
experiment \cite{Zajfman-LiH-2001}. The corresponding
recombination rate $\alpha (E_{c})$ is 
\begin{equation}
\alpha (E_{c})=\sqrt{2E_{c}}\sigma (E_{c})\;.  \label{eq5-Rate}
\end{equation}

This calculated DR rate exhibits an infinite number of resonances near each
closed-channel ionization threshold, associated with autoionizing and
predissociating states of LiH. 
\begin{figure}[tbph]
\begin{center}
\epsfig{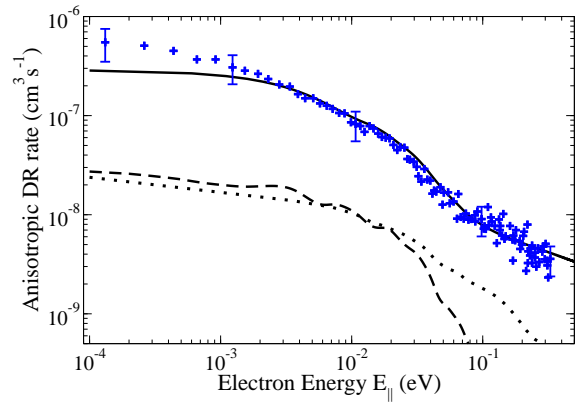}
\end{center}
\vspace{-2mm}
\caption{ DR rate: The solid curve is our calculated, anisotropically-averaged rate for $%
\Delta E_{\Vert }$= 0.1 meV and $\Delta E_{\bot }$ = 12 meV. The broken curve
shows the calculation of \protect\cite{Florescu-Gadea-2004}. The dotted curve
is our truncated result obtained by neglecting the off-diagonal couplings in the
quantum-defect matrix (see the fig.~\protect\ref{fig-sigma-qdm}). The crosses
denote the experimental data \protect\cite{Krohn-thesis}, with a 
few characteristic error bars shown. }
\label{fig-drate}
\end{figure}
To compare with the storage ring experiments \cite{Zajfman-LiH-2001}, 
we must convolve over an anisotropic finite spread in the
electron energy; the spread is different for the
parallel ($\Delta E_{\Vert }=$0.1 meV) and the much broader $\Delta E_{\bot }$ = 12 meV
perpendicular components of the relative velocity vector.
The convolution over parallel and perpendicular energy
distributions has been performed as was outlined in 
\cite{Kokoouline-Greene-2003} and elaborated in
detail in \cite{Kokoouline-Greene-Mosbach}. Figure~\ref{fig-drate}
summarizes our results along with previous experimental and theoretical
results. This figure also demonstrates the results of a numerical test 
conducted to interpret the discrepancy between our
theoretical results and those of  \cite{Florescu-Gadea-2004}. \
Specifically, we have performed one set of calculations that
neglect the off-diagonal $l$-mixing to mimic the calculations performed by 
\cite{Florescu-Gadea-2004}, i.e. using only the diagonal eigenvalue 
form $\mu _{\gamma
}^{\Lambda }(R)$ of quantum-defect matrix (shown in fig.~\ref{fig-sigma-qdm}
c,d). Introduction of this approximation lowers the DR rate by an order of
magnitude, and this artificially restricted calculation agrees with the results of \cite%
{Florescu-Gadea-2004}. \ Thus, the presence of $R$-dependent $l$%
-mixing is crucial for this system. \ Moreover, the rate is predominantly
controlled by $d$-wave collisions, whereas Ref.\cite{Florescu-Gadea-2004}
assumed that it was dominated by $p$-waves.

\begin{figure}[tbph]
\begin{center}
\epsfig{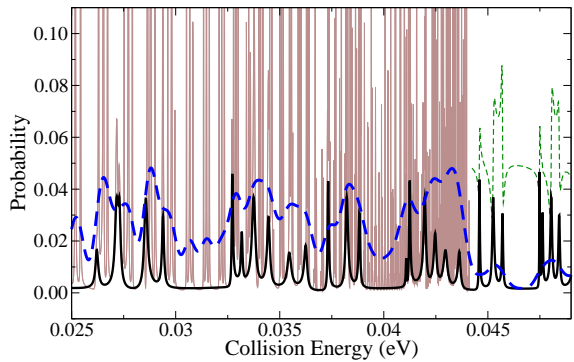}
\end{center}
\vspace{-2mm}
\caption{ Contributions to DR probabilities are shown as functions of 
energy near the first vibrational
threshold: Tht thick broken curve represents the average DR probability. 
The full spectrum of Rydberg
resonances converging to the first threshold is denoted by the 
thin line. Contributions from the second and
higher thresholds are shown as the thick solid curve. 
The thin broken curve is the probability of a vibrational excitation process $%
\protect v  _{0}\rightarrow \protect v  _{1}$, which is only 
energetically allowed above the $v=1$ threshold. }
\label{fig-vrate}
\end{figure}

We now identify the qualitative mechanism
responsible for this high indirect DR rate. Figure~\ref{fig-vrate}
summarizes the probabilities of various DR pathways at an energy near the
first vibrational threshold. For clarity we have neglected rotations in this
qualitative analysis, because we found very little effect of the rotational
frame transformation on the present results at this energy resolution. In
Fig.\ref{fig-vrate}\ no thermal averaging has been applied, and the LiH$^{+}$
ion is initially in the vibrational ground state. 
The DR probabilities shown are the quantity inside the square
brackets in Eq.(\ref{eq5-DR-CS}). The thick broken curve denotes the average
DR probability across the threshold. The DR probability
drops from about 5\% of the incident flux below the threshold down to only
0.6\% above the threshold. The DR probability
below the threshold is built up as the cumulative effect of the dense forest
of Rydberg resonances attached to the first vibrationally excited state,
seen as the thin full curve. The thick full curve shows the contributions to
DR probability when the incident electron is captured into Rydberg states
associated with higher vibrational thresholds. Fig.\ref{fig-vrate} also
explains the reason why the DR flux drops sharply above the $v=1$ threshold.
The thin broken curve shows our calculated probability of vibrational
excitation $0\rightarrow 1$. As can be noticed from the amplitudes for the
probabilities of both processes below and above the threshold, most of the
DR flux just below the threshold turns discontinuously into vibrational
excitation flux once that channel becomes open. These results indicate that
the DR process is controlled by a doorway, namely capture of the incident
electron into a Rydberg state attached to the first vibrational threshold.  
However, if there is no higher-$v$/lower-$n$ perturbing level overlapping the 
total energy of the collision complex, the electron will tend to 
autoionize before DR can take place.  
Throughout the energy range of a multichannel complex resonance, 
though, the initial capture can efficiently pump more energy into vibration
at the first electron recollision, and Fig.\ref{fig-vrate}
shows that this resonant perturbed Rydberg complex 
increases the DR rate by about another factor of 3.  
This complex resonance mechanism for indirect
DR is believed to apply to many other systems that are not controlled by the 
usual simple capture mechanism into a dissociative state.  A hint of its 
importance in H$_3^+$ DR is evident across the energy range 110-160 cm$^{-1}$ 
in Fig.10 of Ref.\cite{KG2004}.

This work was supported in part by NSF grants OISE 0532040 and ITR 0427376. 
We thank S. Tonzani for assistance and discussions in the early stages 
of the project.


\end{document}